\documentclass[manuscript]{acmart}

\usepackage{graphicx}
\usepackage{hyperref}

\AtBeginDocument{%
  \providecommand\BibTeX{{%
    \normalfont B\kern-0.5em{\scshape i\kern-0.25em b}\kern-0.8em\TeX}}}

\copyrightyear{2021} 
\acmYear{2021} 
\setcopyright{rightsretained} 

\acmConference[IUI '21 Companion]{26th International Conference on Intelligent User Interfaces}{April 14--17, 2021}{College Station, TX, USA}
\acmBooktitle{26th International Conference on Intelligent User Interfaces (IUI '21 Companion), April 14--17, 2021, College Station, TX, USA}
\acmDOI{10.1145/3397482.3450733}
\acmISBN{978-1-4503-8018-8/21/04}

\begin{document}

\title[A System for Analyzing Public Sentiments and Discussions about COVID-19]{TweetCOVID: A System for Analyzing Public Sentiments and Discussions about COVID-19 via Twitter Activities}

\author{Jolin Shaynn-Ly Kwan}
\email{jolin\_kwan@mymail.sutd.edu.sg}
\affiliation{%
  \department{Information Systems Technology and Design Pillar}
  \institution{Singapore University of Technology and Design}
  \country{Singapore}
}

\author{Kwan Hui Lim}
\email{kwanhui\_lim@sutd.edu.sg}
\orcid{0000-0002-4569-0901}
\affiliation{%
  \department{Information Systems Technology and Design Pillar}
  \institution{Singapore University of Technology and Design}
  \country{Singapore}
}

\renewcommand{\shortauthors}{Kwan and Lim}

\begin{abstract}
The COVID-19 pandemic has created widespread health and economical impacts, affecting millions around the world. To better understand these impacts, we present the TweetCOVID system that offers the capability to understand the public reactions to the COVID-19 pandemic in terms of their sentiments, emotions, topics of interest and controversial discussions, over a range of time periods and locations, using public tweets. We also present three example use cases that illustrates the usefulness of our proposed TweetCOVID system.
\end{abstract}

\ccsdesc[500]{Human-centered computing~Social networking sites}
\ccsdesc[500]{Information systems~Web and social media search}
\ccsdesc[300]{Information systems~Social networks}

\keywords{COVID-19, Twitter, Sentiment Analysis, Topic Modelling}

\maketitle

\section{Introduction}

The COVID-19 pandemic has infected millions and claimed the lives of hundreds of thousands around the world. In addition, COVID-19 has caused widespread economic impact, resulting in extreme levels of unemployment across all industries. In turn, these medical and economical impacts have created immense emotional stress on the global population. With the ongoing pandemic, it is important to understand the ever-changing impact of COVID-19 as new regulations and developments take place over the world. 

Towards this effort, we present TweetCOVID, a system for understanding the impact of COVID-19 on the general public in terms of sentiments, emotions, topics and controversial discussions. TweetCOVID utilizes publicly available tweets, alongside sentiment analysis (levels of positivity and negativity), emotion detection (anger, fear, sadness, disgust, surprise, anticipation, trust, and joy), topic modelling using Latent dirichlet allocation (LDA), controversial term tracking and various temporal and spatial analysis.

\subsection{Related Work}

Twitter has been used for various applications, such as online misbehaviour detection~\cite{fornacciari2018holistic,chatzakou2017mean}, event tracking~\cite{abdelhaq2013eventweet,lim2018rapid,comito2019bursty}, activity recommendation~\cite{wang2018happiness,madisetty2019event}, etc.
While there has been work using Twitter for COVID-19 related studies, many of these works focus on a single region~\cite{su2020examining,vicari2020covid19} or for a specific single purpose, e.g., misinformation~\cite{gruzd2020going,stephens2020geospatial}. In contrast, our proposed TweetCOVID system provides a comprehensive set of functionalities to study sentiments, emotions, topics, controversial terms across various timeframes and locations. This holistic set of capabilities allow the user to study a flexible range of questions, e.g., what are the public sentiments in a certain country towards the implementation of a lockdown and what are the following public discussions? 

\section{System Architecture and Functionalities}

Our proposed system comprises the main components for data collection/processing, general analytics, sentiment/emotion analysis, topic modelling, controversy tracking and visualisation. These components are further elaborated next.

\begin{figure*}[!ht]
    \centering
    \includegraphics[clip,trim=0mm -10mm 0mm 0mm, width=0.995\textwidth]{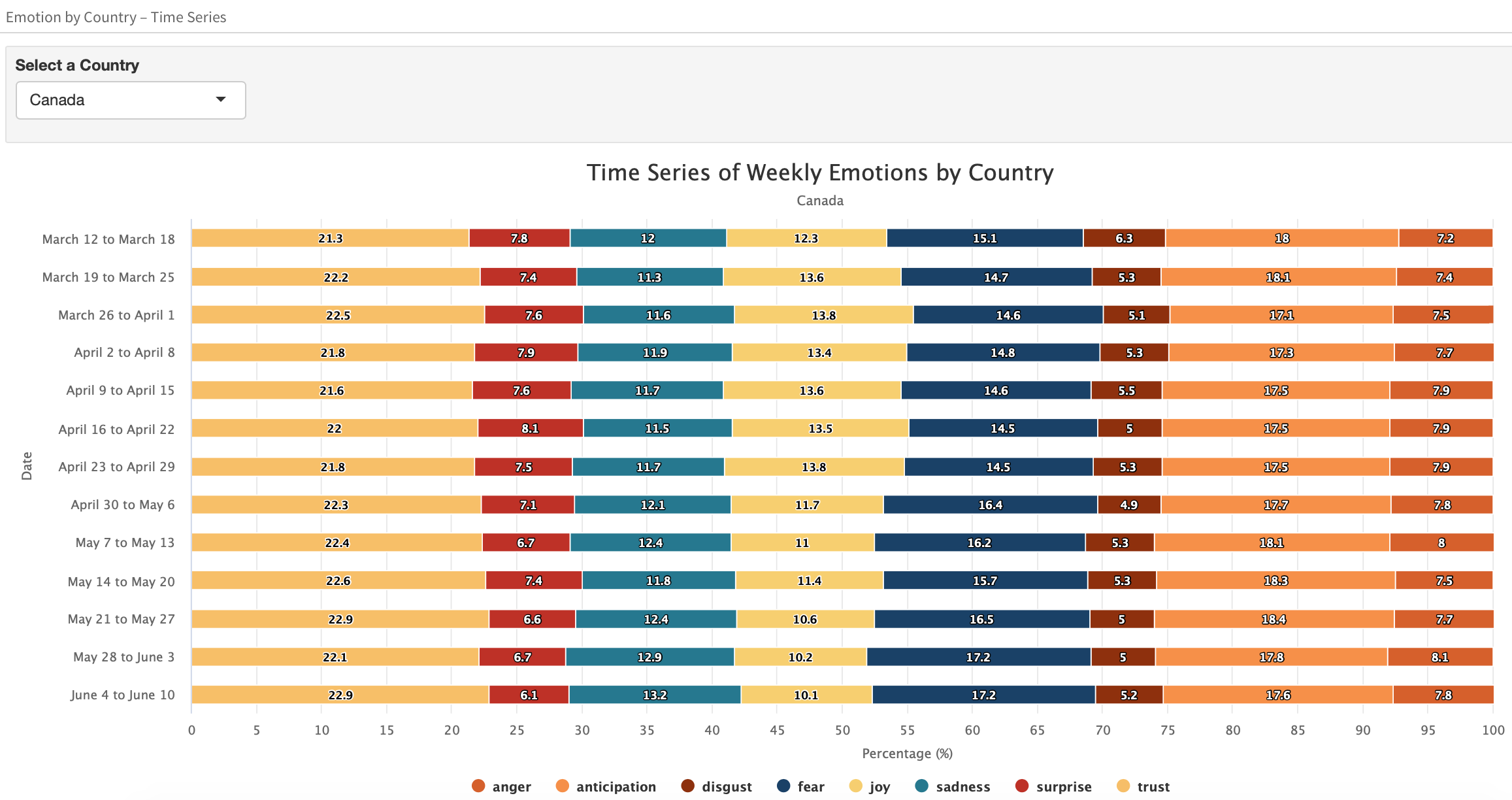}
    \includegraphics[clip,trim=0mm 0mm 0mm 0mm, width=0.995\textwidth]{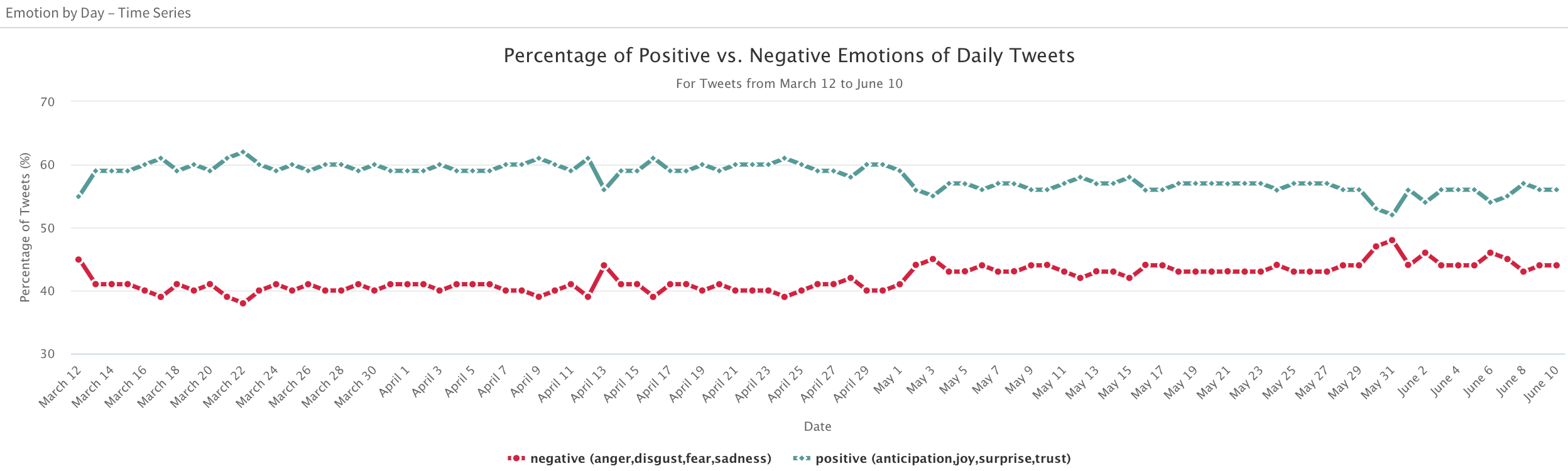}
    \caption{Sample screenshots: Emotions by country and week (top), and overall sentiment over time (bottom).}
    \label{screenshots1}
\end{figure*}

\begin{figure*}[!ht]
    \centering
    \includegraphics[clip,trim=0mm -10mm 0mm 0mm, width=0.995\textwidth]{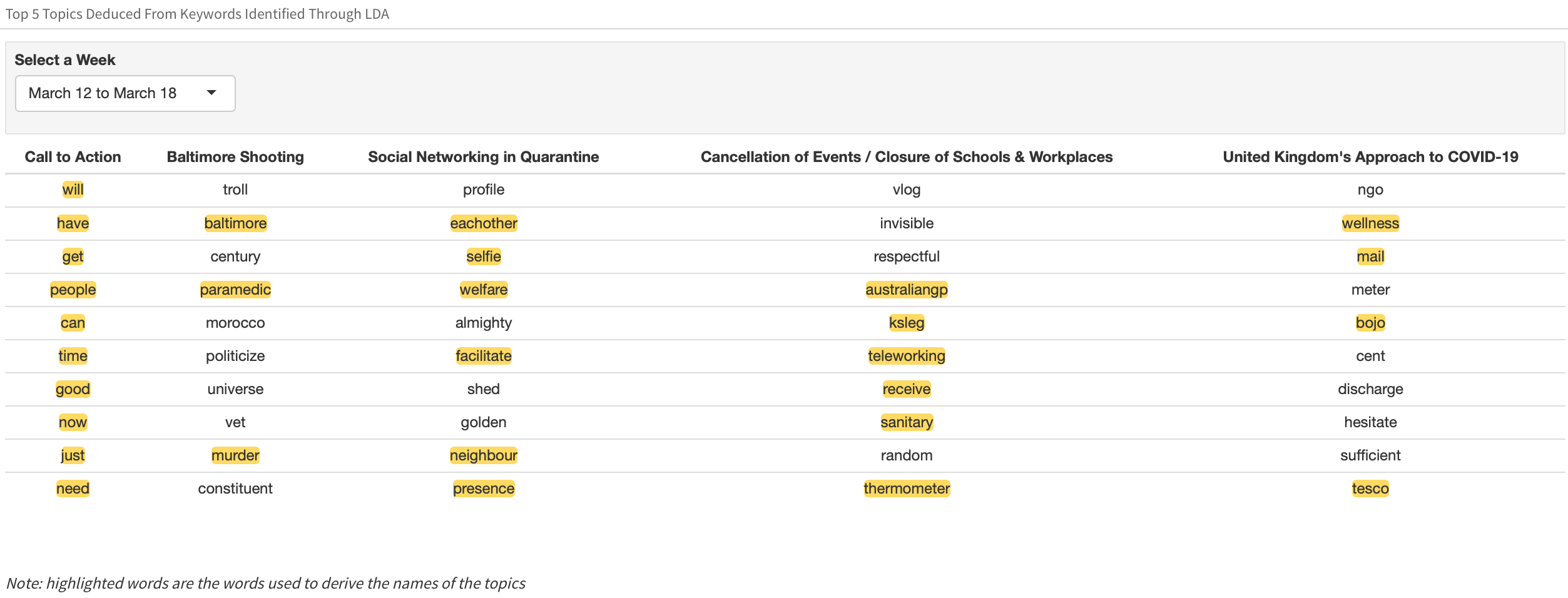}
    \includegraphics[clip,trim=0mm 14mm 0mm 0mm, width=0.995\textwidth]{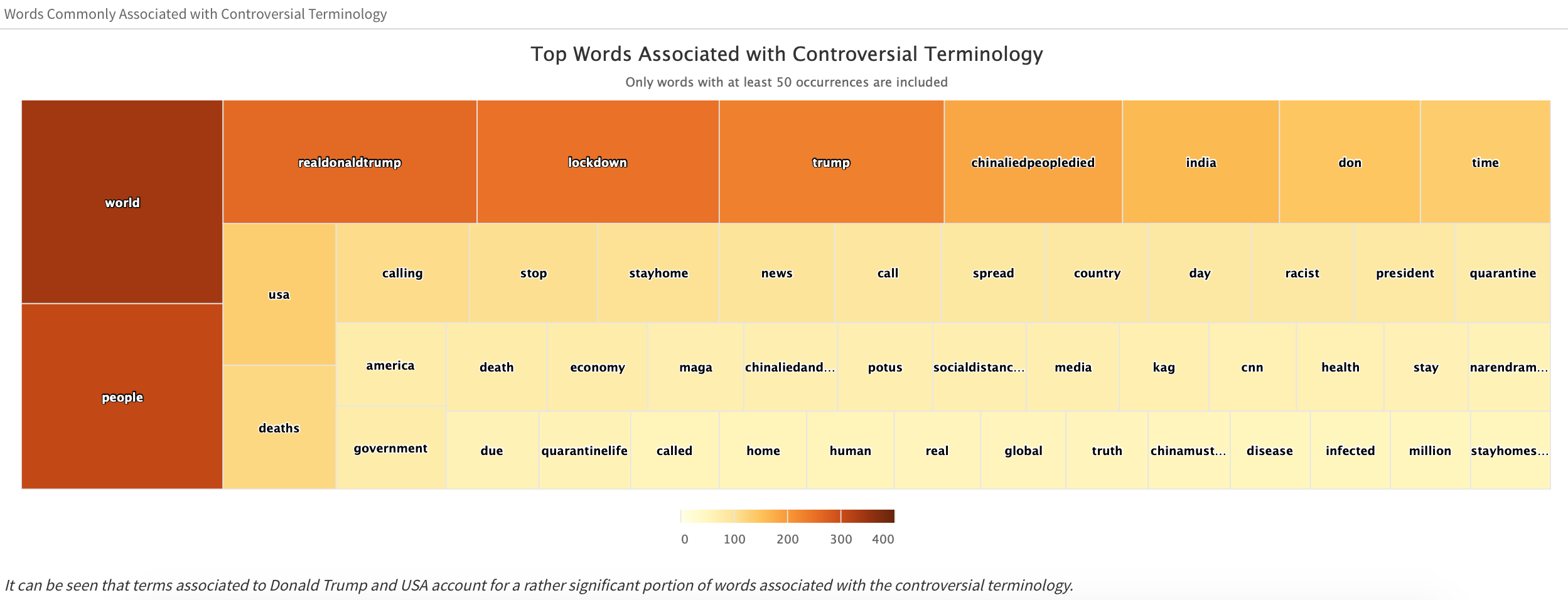}
    \caption{Sample screenshots: Discussion topics by week (top), and top words associated with controversial terms (bottom).}
    \label{screenshots2}
\end{figure*}

\begin{itemize}
    \item {\bf Data Collection/Processing}: This component allows the user to either load in existing Twitter datasets or retrieve datasets via the Twitter API. After loading/collecting these dataset, the standard processing steps of language filtering, tokenization, stemming and lowercase conversion are applied to the dataset before the subsequent analysis.
    \item {\bf General Analytics}: This component shows the general statistics and distribution of the dataset. For example, a user can view the number of tweets posted in specific time periods and/or at geographic locations to understand how frequent the discussion about COVID-19 is in specific areas.
    \item {\bf Sentiment/Emotion Analysis}: This component shows the general sentiments and specific emotions associated with COVID-19 related discussions. We use the AFINN lexicon~\cite{nielsen2011new} for sentiment analysis that allows us to determine the positivity and negativity of specific tweets, which we further aggregate for the overall COVID-19 discussions. For specific emotions, we use the NRC Word-Emotion Association Lexicon~\cite{mohammad2013nrc} that allows us to determine the extent which a tweet displays the eight primary emotions based on Plutchik theory~\cite{plutchik1980general}. Figure~\ref{screenshots1} shows two screenshots of the sentiment and emotion analysis components.
    \item {\bf Topic Modelling}: This component displays the various topics that were discussed over time during the course of the pandemic. We used the popular LDA~\cite{blei2003latent} on the set of weekly tweets to determine the main topics discussed on a weekly basis. LDA represents tweets as a random mixture over latent topics and each topic is in turn represented by a set of words. Figure~\ref{screenshots2} (top) shows a screenshot of this topic modelling component.
    \item {\bf Controversy Tracking}: This component tracks and identifies tweets that include various controversial terms, such as "wuhan virus", "chinese virus", "kung flu", etc. These tweets are controversial in their content and designed to incite hatred towards a specific group of people. Figure~\ref{screenshots2} (bottom) shows a screenshot of this controversy tracking component.
    \item {\bf Visualisation}: This component provides a wide range of visualisation techniques to best display the processed results from the above-mentioned components. In addition, a user is able to drill-down and present these analysis across different time periods and locations.
\end{itemize}
    
Figures~\ref{screenshots1} and~\ref{screenshots2} shows various example screenshots of our TweetCOVID system. For a more interactive demonstration of these and additional features, we refer interested readers to \url{https://youtu.be/ut2EzLeuFwo} for a video demonstration.

\section{Example Use Cases / Demonstration}

Using the set of functionalities provided by TweetCOVID, we can utilize the system for various use cases, which we describe with three examples. For a more detailed discussion of the main findings, we refer interested readers to our earlier paper~\cite{kwan2020covid} that elaborates on various research questions and findings related to the COVID-19 pandemic.

{\bf Use Case 1: Understanding public sentiments and emotions in relation to global events}. Using our system, we observe that public sentiments were noticeably more negative on 12 March 2020 as compared to the weeks before and after. Further analysis shows that anxiety and fear levels were heightened after the WHO Director-General declared COVID-19 as a pandemic on 11 March 2020. Thereafter, there was a corresponding increase in discussions relating to the pandemic and the potential impact on the users' daily life.

{\bf Use Case 2: Detecting adjacent topics during the course of the pandemic}. Apart from the COVID-19 pandemic, there were various global incidents that happened concurrently with the pandemic. Our TweetCOVID system was able to pick up such incidents in the form of discussion topics. For example, we detected the topic of the Baltimore Shooting over 12 to 18 March 2020 and how certain users were linking this incident back to COVID-19.

{\bf Use Case 3: Understanding the use of controversial terms during the pandemic}.
Our proposed system also tracks the use of controversial terms, such as "wuhan virus", "chinese virus", "kung flu", and the context they were used in. We found that more than half of these usage were from the USA, with a majority from political figures during March 2020. Additionally, the racist nature of the controversial terms can be seen through the co-occurrence of terms such as “chinaliedpeopledied”, “racist” and “chinamustexplain”. These terms suggest misplaced blame upon people of Asian descent for the pandemic and further perpetuates racism in countries.

\section{Conclusion}
In this paper, we present a demonstration of the TweetCOVID system for understanding the impact of COVID-19 on the general public using publicly available tweets. TweetCOVID offers a range of functionalities, including data collection/processing, general analytics, sentiment/emotion analysis, topic modelling, controversy tracking and visualisation. We also demonstrate three use cases that our system can be used for, including understanding sentiments/emotions, detecting topics alongside COVID-19 and understanding controversial terms used. Future work will include exploring other topic models that are relevant for short text~\cite{jonsson2015evaluation}.

\begin{acks}
This research is funded in part by the Singapore University of Technology and Design under grant SRG-ISTD-2018-140.
\end{acks}

\bibliographystyle{ACM-Reference-Format}
\bibliography{covidDemo}

\end{document}